# Correlating Structural, Electronic, and Magnetic Properties of Epitaxial VSe$_2$ Thin Films

Guannan Chen[1], Sean T. Howard[1], Aniceto B. Maghirang III[2], Kien Nguyen Cong[3], Kehan Cai[4], Somesh C. Ganguli[5], Waclaw Sweich[6], Emilia Morosan[7,8], Ivan I. Oleynik[3], Feng-Chuan Chuang[2], Hsin Lin[9], Vidya Madhavan[1]

[1] Department of Physics and Frederick Seitz Materials Research Laboratory, University of Illinois Urbana-Champaign, Urbana, Illinois 61801, USA
[2] Department of Physics, National Sun Yat-sen University, Kaohsiung, Taiwan
[3] Department of Physics, University of South Florida, Tampa, Florida 33620, United States
[4] Department of Chemistry, Princeton University, Princeton, New Jersey 08544, United States
[5] Department of Applied Physics, Aalto University School of Science, PO Box 11100, 00076 Aalto, Finland
[6] Departments of Materials Science and Engineering, University of Illinois at Urbana-Champaign, Urbana, IL 61801, USA
[7] Department of Physics and Astronomy, Rice University, Houston, Texas 77005, USA
[8] Rice Center for Quantum Materials, Rice University, Houston, Texas 77005, USA
[9] Institute of Physics, Academia Sinica, Taipei, Taiwan

## Abstract

The electronic and magnetic properties of transition metal dichalcogenides are known to be extremely sensitive to their structure. In this paper we study the effect of structure on the electronic and magnetic properties of mono- and bilayer VSe$_2$ films grown using molecular beam epitaxy. VSe$_2$ has recently attracted much attention due to reports of emergent ferromagnetism in the 2D limit. To understand this important compound, high quality 1T and distorted 1T films were grown at temperatures of 200℃ and 450℃ respectively and studied using 4K Scanning Tunneling Microscopy/Spectroscopy. The measured density of states and the charge density wave (CDW) patterns were compared to band structure and phonon dispersion calculations. Films in the 1T phase reveal different CDW patterns in the first layer compared to the second. Interestingly, we find the second layer of the 1T-film shows a CDW pattern with 4a × 4a periodicity which is the 2D version of the bulk CDW observed in this compound. Our phonon dispersion calculations confirm the presence of a soft phonon at the correct wavevector that leads to this CDW. In contrast, the first layer of distorted 1T phase films shows a strong stripe feature with varying periodicities, while the second layer displays no observable CDW pattern. Finally, we find that the monolayer 1T VSe$_2$ film is weakly ferromagnetic, with ~3.5 $\mu_B$ per unit similar to previous reports.

# Introduction

Two dimensional (2D) materials such as graphene have attracted much recent attention due to novel electronic and physical properties that accompany reduced dimensionality. While graphene has a large range of potential applications, the lack of an electronic band gap limits its use in optical and semiconducting devices [1]. Another 2D material system of interest is transition metal dichalcogenides (TMDs). TMDs are layered materials containing two chalcogen atoms per transition metal atom, displaying strong intra-layer bonding and weak van der Waals inter-layer bonding. The weak inter-layer bonding of TMDs facilitates control of film thickness via growth or exfoliation down to sub-monolayer. As TMDs are reduced to 2D, novel physics often emerges such as itinerant magnetism [2], an indirect to direct bandgap transition [3, 4], quantum spin Hall effect [5], and strongly enhanced charge density wave (CDW) order [6]. This breadth of phenomena makes 2D TMDs a promising platform both for the development of next generation devices and important fundamental studies.

$VSe_2$ has recently received enormous research interest due to reports of emergent room temperature ferromagnetism in the 2D limit. This however remains controversial as several theoretical [7, 8, 9, 10] and experimental [11, 12, 13] studies both confirm and deny the possibility of a ferromagnetic phase in this compound. The films are also interesting due to the variety of CDW patterns observed which are distinct from the bulk sample [14, 15], raising questions about the role of Fermi surface nesting and phonons in CDW formation. Bulk $VSe_2$ is paramagnetic [16, 17, 18] and noteworthy for being one of the few materials exhibiting a three-dimensional CDW (4a x 4a x 3c) [19]. The bulk material has been shown to be stable in 1T octahedral structure (Fig. 1a and 1b). In general, however, TMDs can also be found in 2H trigonal prismatic structure and distorted 1T structure (Fig. 1c and 1d), including $1T_d$ orthorhombic and 1T' monoclinic, which occurs when the chalcogen atoms in the 1T phase dimerize [1, 20]. Most reports of 2D $VSe_2$ films have been carried out on the 1T phase, leaving the synthesis and electronic properties of other polymorphs largely unexplored.

In this paper we report the molecular beam epitaxy (MBE) growth of $VSe_2$ films on bilayer graphene (BLG)/6H-SiC (0001) substrates at two different growth temperatures, resulting in two distinct polymorphs. The synthesis procedures were characterized with reflection high-energy electron diffraction (RHEED) and scanning tunneling microscopy (STM) to discern structural and electronic properties compared to bulk $VSe_2$ crystals. At lower growth temperature of 200 °C, the usual 1T phase is grown, which exhibits a unique incommensurate CDW in the monolayer in contrast to previous reports [11, 13, 14, 21, 22]. Interestingly, the CDW in bilayer is a two-dimensional projection of the commensurate bulk CDW structure. For the higher growth temperature of 450 °C, a striped discommensurate CDW in the monolayer is revealed which is a precursor to the distorted 1T phase (Fig. 1c and 1d) observed in the bilayer. Our magnetization measurements show that the monolayer $VSe_2$ samples grown at low temperature are weakly ferromagnetic with Curie temperature higher than 300 K. Intriguingly, the 1.5-layer VSe$_2$ samples grown at higher temperature also exhibit ferromagnetic behavior at room temperature or below, but with a lower magnetic moment per V atom.

## Methods

VSe$_2$ thin films were grown on bilayer graphene (BLG) on SiC using a home-built MBE system with a base pressure $< 1 \times 10^{-9}$ Torr. For BLG growth, 6H-SiC (0001) substrates were washed in acetone and isopropanol and then loaded into MBE chamber. The substrates were degassed at 650 °C for 2 to 3 hours, then flash annealed 45 times between 650 °C and 1300 °C. High purity V (99.8%) and Se (99.999%) were evaporated from an e-beam evaporator and a dual-filament low temperature Knudsen cell respectively. The fluxes of V and Se were measured by quartz crystal monitor, with the flux ratio kept between 1:20 to 1: 30. The growth processes were monitored by in-situ RHEED. The thin films grown at 200 °C and 450 °C are labelled low growth temperature (LGT) samples and high growth temperature (HGT) samples respectively.

After growth, the samples were transferred to a low temperature scanning tunneling microscope (STM) using a home-built "vacuum suitcase" to prevent the degradation of the sample quality. The vacuum during the transfer was less than $1 \times 10^{-9}$ Torr. STM/S measurements were performed at 4K. In the STM measurements, electrochemically etched and vacuum annealed tungsten tips were used. For comparison, bulk VSe$_2$ single crystals were cleaved in-situ at a base pressure $< 2 \times 10^{-9}$ Torr and transferred into the same low temperature STM.

Before the magnetic property measurements, samples were capped with 10nm of amorphous Se immediately after growth. The samples were then taken out of the MBE chamber and mounted into Quantum Design Magnetic Properties Measurement System (MPMS). The magnetization M-H curves were measured in Superconducting Quantum Interference Device (SQUID) Vibrating Sample Magnetometer (VSM) mode with in-plane magnetic field. Magnetic moment per formula unit is roughly estimated by dividing the saturation magnetization, converted to Bohr magnetons, by the number of formula units which is obtained from the surface area, thickness and volume of VSe$_2$ unit cell.

The first-principles calculations were carried out using the Vienna Ab initio Simulation Package (VASP) [23, 24] with the projected augmented wave (PAW) [25] potentials. The exchange–correlation functional was treated within the Perdew–Burke–Ernzerhof (PBE) generalized gradient approximations (GGA) [26-30]. The cut-off energy used throughout the calculations was set to 400 eV. Atomic positions were optimized for each lattice constant value considered until the residual forces were no greater than $10^{-3}$ eV/Å. The criteria for energy convergence for self-consistency was set at $10^{-6}$ eV. The vacuum region along the z direction was set to approximately 15 Å to prevent interactions between the repeated monolayer/bilayer slabs under periodic boundary condition. A Γ-centered Monkhorst-Pack [31] grid of 12×12×1 in the first Brillouin zone was used for calculating atomic structures and lattice relaxations. However, a denser grid of 36×36×1 was used for density of states calculations.

To investigate lattice dynamics, phonon dispersion is calculated using supercell method as implemented in Phonopy code [32]. 4×4 supercells and 1×1 cells are considered for phonon calculations of 1T and 4×4 CDW structures, respectively. Crystal structure of 4×4 CDW is determined by displaying atoms of 4×4 perfect 1T supercell along the eigen-vector of a soft mode at a commensurate q-point followed by atomic relaxation in the fixed supercell.

## Results and Discussion

Previous works on $VSe_2$ thin films report several distinct CDW patterns, as well as existence and absence of ferromagnetism, which highlights the sensitivity of resultant film properties to growth parameters and substrate choices [11, 13, 14, 15]. The exploration of the substrate and synthesis parameter space is therefore important in fine tuning film properties. BLG was chosen as a substrate in hopes of approximating a free-standing film for two reasons. First, as graphene does not have a fermi surface near the Γ point where $VSe_2$ has a hole-like band, substrate-film interaction could be minimized. Second, a large lattice mismatch between graphene and Se lattice encourages weak van der Waals bonding between BLG and the film. However, previous studies report a dependence of the CDW phase in $VSe_2$ thin films with relative substrate-film angle. Therefore, heterostructure effects cannot be totally neglected [14].

Two different synthesis conditions were used in our experiments; LGT, where the substrate was held at 200 °C, and HGT, where the substrate was held at 450 °C. The BLG grown on SiC is atomically flat, observed by RHEED whose pattern is shown in Fig. 2a. High quality 2D growth is confirmed with sharp streaks in RHEED images (Fig. 2b), which are seen for the 0.5 layer at both growth temperatures. A growth rate of 0.06 layer/minute allowed for control of film thickness. Since our studies focused on the first and second layer, we aimed for 1.5-layer film growth in both conditions. Both growth procedures produced large terraces of $VSe_2$ on BLG (Fig. 2d and 2f).

The 1.5-layer LGT film shows the same RHEED pattern as 0.5-layer as shown in Fig. 2c. The absence of graphene RHEED pattern means the film fully covers the substrate. Performing STM on the 1.5-layer LGT film, we image a triangular Se lattice with lattice constant a=0.34 nm (Fig. 3a and 3d), which is consistent the bulk lattice constant as well as with other reports of $VSe_2$ films [19, 33]. In the region where multiple layers (including a small exposed BLG area) are visible, a linecut profile across the edges is shown in Fig. 2f (the red line in Fig. 2e). A step height of 0.6 nm is seen between the first and second layer VSe$_2$, consistent with the c-axis lattice constant in bulk of 0.61nm [33]. On the other hand, the height difference between BLG and first layer is 0.8 nm, which is slightly larger than the lattice constant c in the bulk VSe$_2$. This is consistent with previous experimental reports [34, 35] and can be identified as monolayer since 0.8 nm is significantly closer to monolayer (0.61 nm) thickness than the 2-layer-thickness (1.22 nm). This difference in heights is likely due to interfacial effects. Atomic resolution STM images on the monolayer also show several 2D CDW patterns in different areas (Supplemental Fig. 1). The most frequently observed CDW is shown in Fig. 3a with CDW vectors of $q_1 = 4.2a$ by $q_2 = 4.6a$ approximately, concomitant with a strong incommensurate 1D stripe order of periodicity $d = 0.79$ nm or 2.33a. While the 1D order occurs 33 degrees deviated from a Se lattice direction, the 2D CDW order forms an oblique lattice with an angle of around 114 degrees between the two CDW lattice vectors. Although the different 2D CDW patterns have minor variances in the angle and magnitude, the 1D stripe feature is observed in all monolayer scans with the same angle and periodicity. The STM spectra present on the LGT monolayer show a large suppression of the density of states near the Fermi energy, with the suppression beginning at -30 meV, ending at 18 meV, and a minimum at the Fermi energy. While the differential conductance at the zero bias is significantly suppressed, it is non-zero as seen in Fig. 3c.

The LGT bilayer sample resembles the bulk sample (Supplemental Fig. 2a), displaying a commensurate 2D CDW pattern with 4a × 4a periodicity (Fig. 3d). Our phonon dispersion calculations (Fig. 3e) display three noticeable imaginary modes at q$_1$, q$_2$ and q$_3$. The mode at $q_1 \sim$ -1/2 $\overline{\Gamma M}$ has the largest negative frequency, and the displacing atom along the eigenvector of this mode results in a 4a × 4a commensurate CDW structure. According to our calculation, the energy of this CDW state is 11 meV/atom lower than that of the normal state. Thus, the 4a × 4a CDW is indeed preferred in this system and the electron-phonon interaction plays an important role in its formation. In the STS measurements (blue curve in Fig. 3f), a large peak in the density of states around -20 meV is observed. A suppression of the density of states near the Fermi level is seen in the bilayer, like in the monolayer LGT samples. We also performed the first-principles calculation for the density of states curve. The DOS curve of bilayer 1T-VSe$_2$ is shown as the red curve in Fig. 3f. The calculation fits well with the experimental data regarding the positions of features. The peaks near Fermi energy indicate the 3d band of V atoms and the split of peaks results from the duplication of electron bands in a bilayer system. In the supplemental material, we further compare the spectrum of the second layer with the one in the bulk sample (Supplemental Fig. 2b). A similar peak feature appears in both spectra but is sharper in the bilayer, which confirms that it is the V 3d band. Comparison of the bilayer to the monolayer in the LGT films demonstrates the significant impact reduced dimensionality has on just one additional layer of growth.

Remarkably, by only changing the substrate growth temperature by a few hundred degrees, the HGT films exhibit drastically different properties. While RHEED images of the monolayer look identical to the LGT (Fig. 2b), STM images show strong 1D stripe feature with varying periodicity.

As seen in Fig. 4a, the stripes occur along a lattice direction with periodicities of either 4 atoms, 5 atoms, or 6 atoms. Along the direction of stripes, the lattice constant, 0.34 nm, is same as the one in LGT samples. Taking a linecut along the stripe direction (Fig. 4e), it's clear that the distances between two adjacent atoms are uniform. However, along the other two lattice directions (the height profile shown in Fig. 4f), it appears that the atoms are not evenly spaced and the average atomic separation, 0.31 nm, is almost 10% smaller than the one along the stripes and in LGT samples. To confirm the universality, we took topographies in different areas with stripes along various directions. We found that the difference between lattice constants along and cross the stripe direction is consistent. Interestingly, the spectra on bright and dark stripes are different (Supplemental Fig. 3) which suggests an electronic origin (such as a CDW) for the stripes. 1D CDWs with varying periodicities have been explained using the theory of discommensurations, where slightly incommensurate CDWs become commensurate over a region and undergo a phase slip between regions to lower their energy [36]. Discommensurate CDWs have been previously observed in bulk $NbSe_2$, caused by local strain [37]. In our HGT $VSe_2$ samples, we speculate that strain induced by synthesis conditions or growth specific heterostructure interactions may cause dimerization and discommensurate CDW in the monolayer. We note that a discommensurate 1D CDW has not been observed in previous $VSe_2$ monolayer studies.

Comparing the spectra on the LGT monolayer to the HGT monolayer, we find a smaller reduction of the spectroscopic density of states near the Fermi energy and broader energy range over which this reduction occurs. As shown in Fig. 4b there are two well defined peaks in the spectra, occurring at -56 meV and + 40 meV. The smaller reduction could be attributed to increasingly imperfect Fermi surface nesting caused by strain.

Accompanying HGT bilayer growth are additional streaks in the RHEED image, as shown in Fig. 2d. These streaks occur halfway between the center spot and the most prominent RHEED streaks, implying the onset of a 2a structural periodicity. STM images of the HGT bilayer show a one-dimensional pattern of bright-dark-bright-dark atoms along a lattice direction (Fig. 4c). The RHEED pattern, STM images, and previous reports on 1T' TMDs [5, 38] lead us to conclude that the bilayer HGT film grows in the distorted 1T phase. As illustrated in Fig. 1c and 1d, the V atoms in the distorted 1T phase dimerize, causing the bond lengths of neighboring Se atoms to change. Se atoms in the center of a V dimer are slightly elevated in the inter-layer direction and Se atoms in-between dimers are slightly depressed. This elevation and depression are responsible for the contrast seen in the STM image and is obviously exhibited in a linecut profile Fig. 4g. As shown in Supplemental Fig. 4, the spectra are almost identical across the stripes, consistent with our RHEED analysis that the stripes result from lattice distortion. Since the distorted film thickness is just one layer, there is no need to distinguish between $1T_d$ and 1T' phase here. The difference between these two phases is the stacking method, with the 1T' phase being monoclinic and the $1T_d$ phase being orthorhombic.

Though the structures of monolayer and bilayer are remarkably different, their spectra are qualitatively similar. With positive sample bias, there are three peaks at approximately same positions as the spectra of the first layer, though the intensities are different. But below the Fermi energy, the density of states curve in the second layer is very different from the one in the monolayer. Despite the bilayer not having CDW order, the spectrum also features a reduction of the density of states near the Fermi energy. We also carried out first-principles calculation for 1T' and $1T_d$ phase bilayer $VSe_2$. The comparison between experiment and calculation are shown in Supplemental Fig. 5. We find that unlike the calculations for the 1T films, the agreement between theory and experiment is not very good. We attribute this to the fact that distorted 1T samples grow on a layer of strained 1T $VSe_2$, which is not same as the calculation condition.

Beside the electronic and structural properties of $VSe_2$ films in the atomic limit, the magnetic properties are also of significant interest. According to the M-H curves from MPMS measurements of our samples at different temperatures, the monolayer LGT $VSe_2$ sample is ferromagnetic, with ~3.5 $\mu_B$ per formula unit similar to previous reports [14] and the Curie temperature is higher than 300 K. Intriguingly, the 1.5-layer HGT $VSe_2$ samples also exhibit a ferromagnetic behavior at room temperature or below, but with a weaker magnetic moment, ~1.3 $\mu_B$ per formula unit. More details are displayed in Supplemental Fig. 6.

## Conclusion

In this paper we report the growth of $VSe_2$ films on BLG/SiC at 200 °C and 450 °C, labelled LGT and HGT respectively. The LGT films show 2D incommensurate CDW patterns in the monolayer, with a periodicity that is different from those of previous reports. This evolves into a commensurate 4a × 4a CDW in the bilayer that is similar to the bulk. The HGT films show new phenomenology when compared to the LGT films and previous reports. The HGT monolayer displays a 1D discommensurate CDW, indicative of strain. The HGT bilayer grows in the distorted 1T phase, which is so far rarely reported in $VSe_2$ and is likely stabilized by interactions with the monolayer. Our results on the HGT films allow for the study of two new phenomena in $VSe_2$ films: a discommensurate 1D CDW and the distorted 1T phase. Specifically, the distorted 1T phase provides more possibilities for synthesizing different phases of TMDC thin films and realizing new heterostructures with potentially exotic properties.

## Acknowledgements

V.M. and E.M. gratefully acknowledge support from NSF DMREF 1629068 and NSF DMREF 1629374 for STM studies and sample characterization. MBE growth and characterization was made possible by support from the Gordon and Betty Moore Foundation's EPiQS Initiative through Grant GBMF4860. F.C.C. acknowledges support from the National Center for Theoretical Sciences and the Ministry of Science and Technology of Taiwan under Grants No. MOST-107-2628-M-110-001-MY3. He is also grateful to the National Center for High-performance Computing for computer time and facilities.

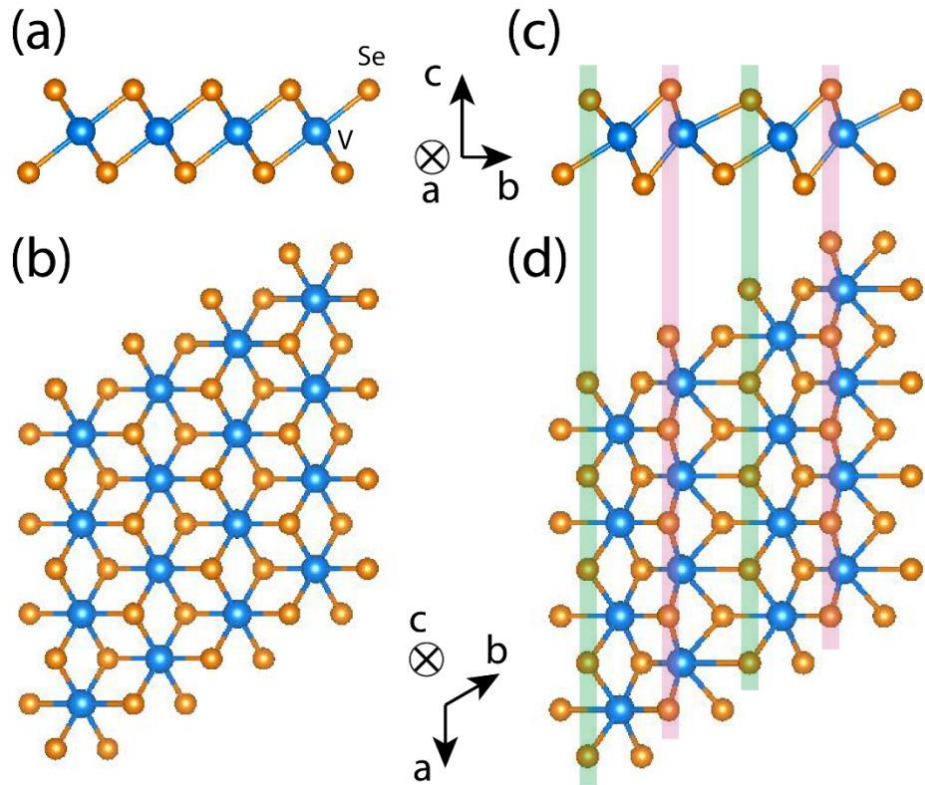

**Figure 1| Structure Schematic of 1T and distorted 1T VSe₂.** The blue circles represent V atoms and the orange circles represent Se atoms. **a,** side view and **b,** top view of 1T-VSe₂. **c,** side view and **d,** top view of distorted 1T-VSe₂. The green and pink ribbons highlight the lower and higher top Se atoms correspondingly.

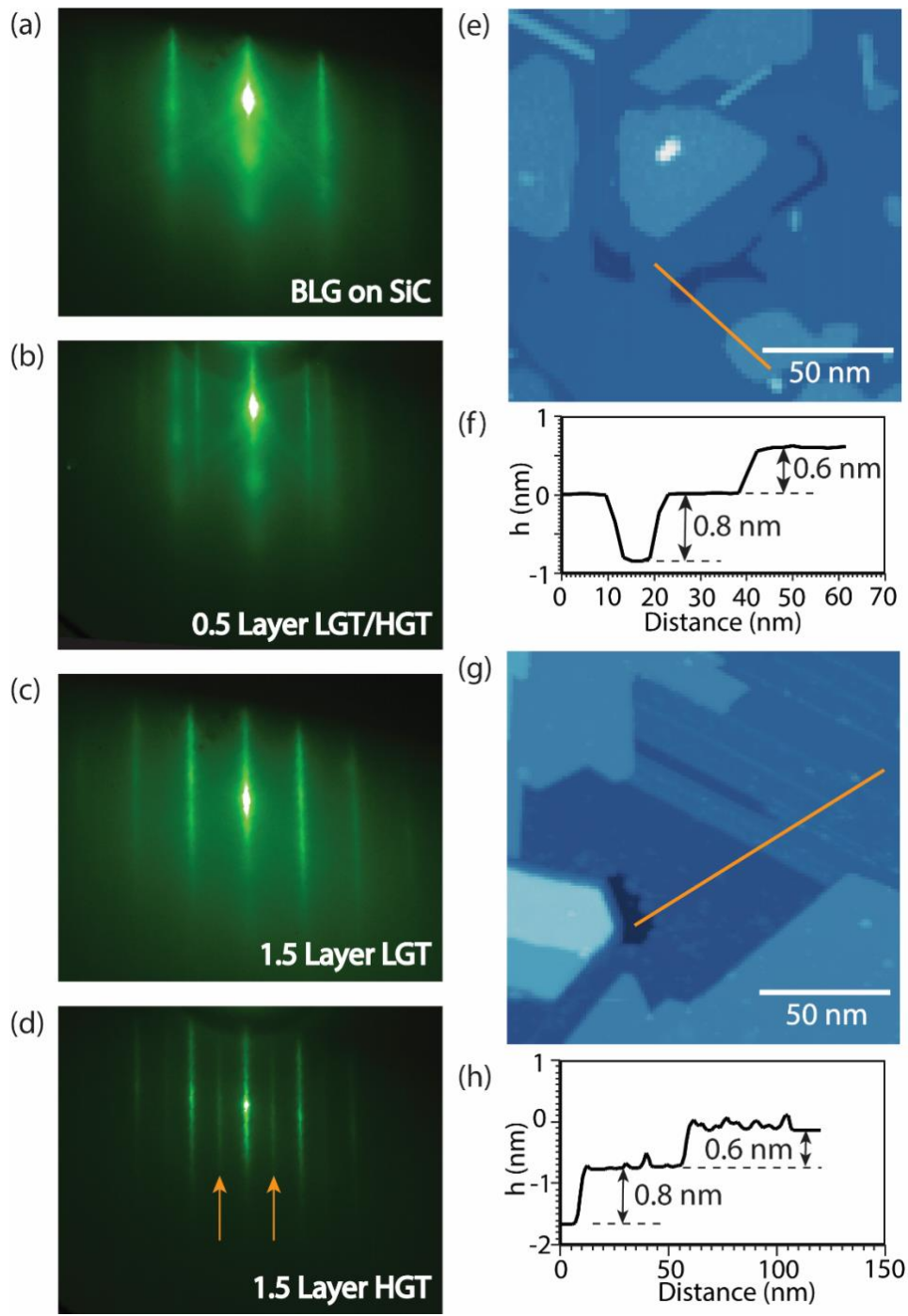

**Figure 2| Characterization of epitaxially grown VSe₂ films by RHEED and STM. a, b, c, d,** RHEED pattern of BLG grown on 6H-SiC (0001) substrate, LGT and HGT films after 0.5 layer growth, LGT film after 1.5 layer deposition and HGT film after 1.5 layer deposition. In **d**, the extra streaks are marked by the orange arrows, which indicate the 1T' phase film forms. **e, g,** large STM topographies of 1.5-layer LGT and 1.5-layer HGT samples (150 nm × 150 nm). **f, h,** height profiles of the orange lines in **e** and **g**, respectively.

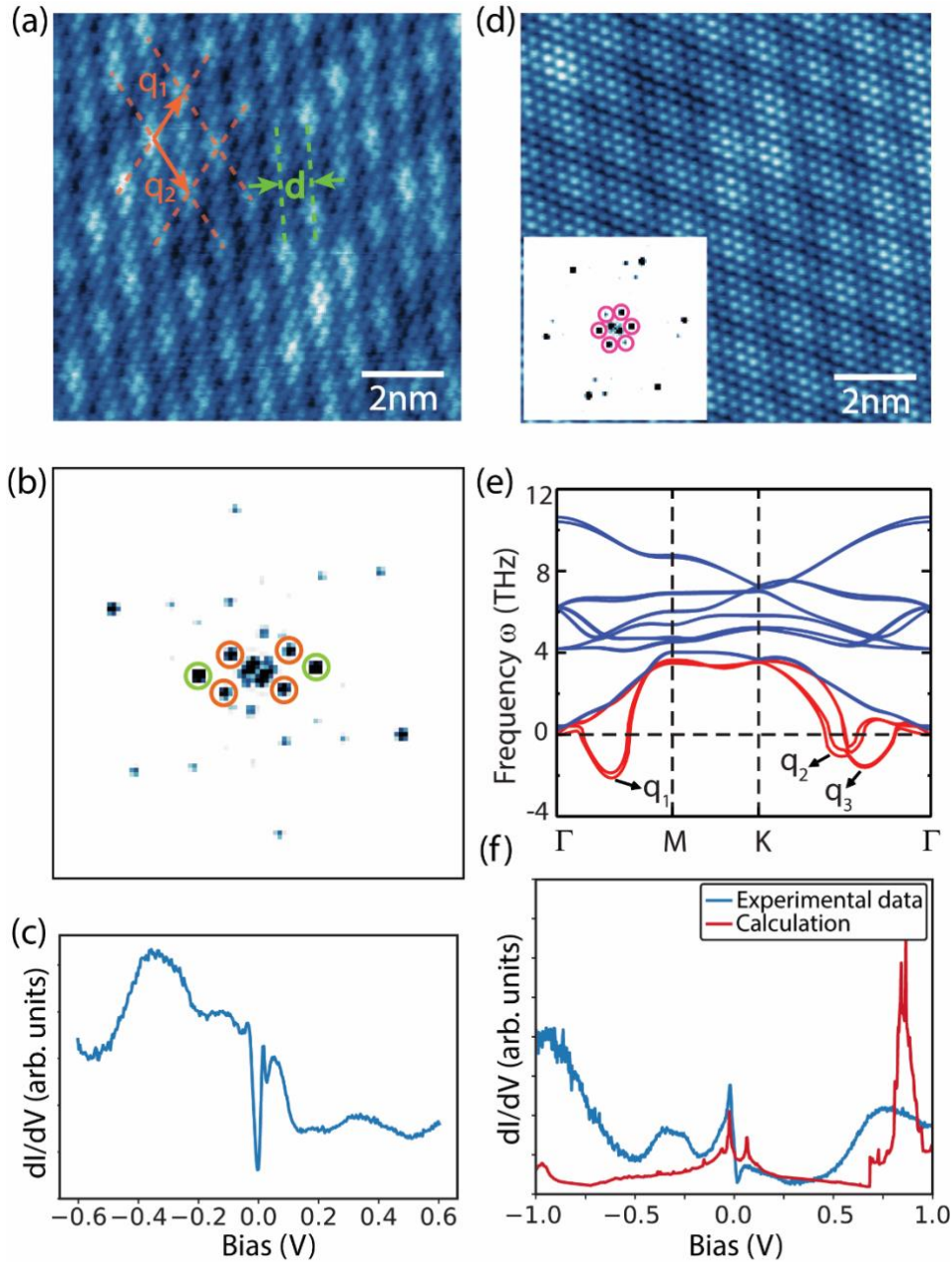

**Figure 3| STM analysis of LGT VSe₂ film. a,** atomic resolution STM topography of the 1st layer in LGT sample (10 nm × 10 nm). The 2D CDW vectors are shown as the orange arrows. And green lines indicate the 1D stripe order with periodicity d. **b**, 2D Fast Fourier Transform (FFT) of **a**. The orange and green circles represent the 2D CDW order and 1D stripe order, respectively. **c**, the typical dI/dV spectrum of the 1st layer LGT. **d**, the atomic resolution STM topography of the 2nd layer in LGT sample (10 nm × 10 nm). Inset is the 2D FFT. The pink circles in the FFT indicate the peaks associated with 4a × 4a CDW. **e**, calculated phonon dispersion relations for bilayer VSe₂. **f**, comparison of dI/dV spectra of the 2nd layer of LGT sample (blue line) and density of state curve from first-principles calculation (red line).

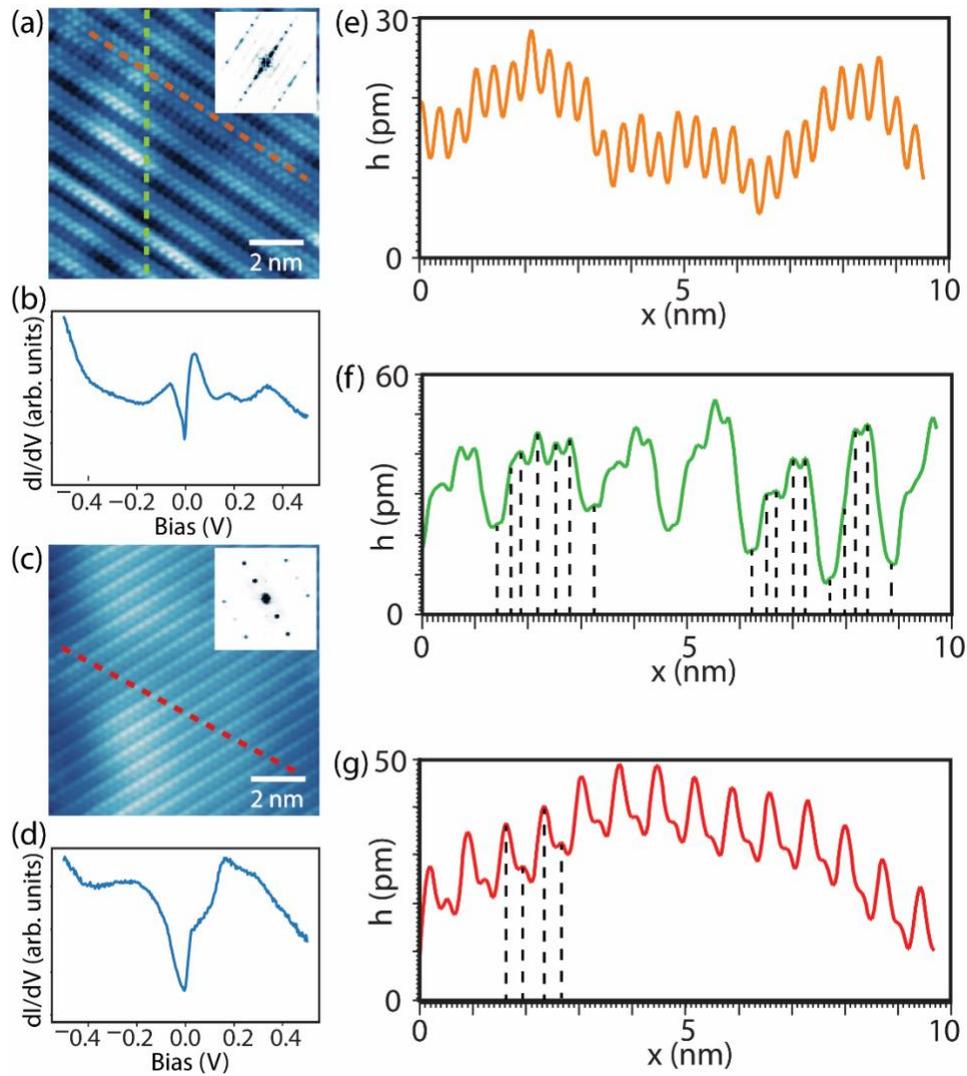

**Figure 4| STM analysis of HGT VSe$_2$ film. a, c,** the STM topographic images of the 1$^{st}$ layer and 2$^{nd}$ layer of HGT samples (10 nm × 10 nm). The insets are the corresponding 2D FFT images. **b, d,** the typical dI/dV curves measured on the 1$^{st}$ layer and 2$^{nd}$ layer of HGT VSe$_2$ films respectively. **e, f, g,** the height profiles of the orange and green line in **a** and red line in **c**, respectively. In **f** and **g**, the black dash lines show the positions of atoms.

# Supplemental Material

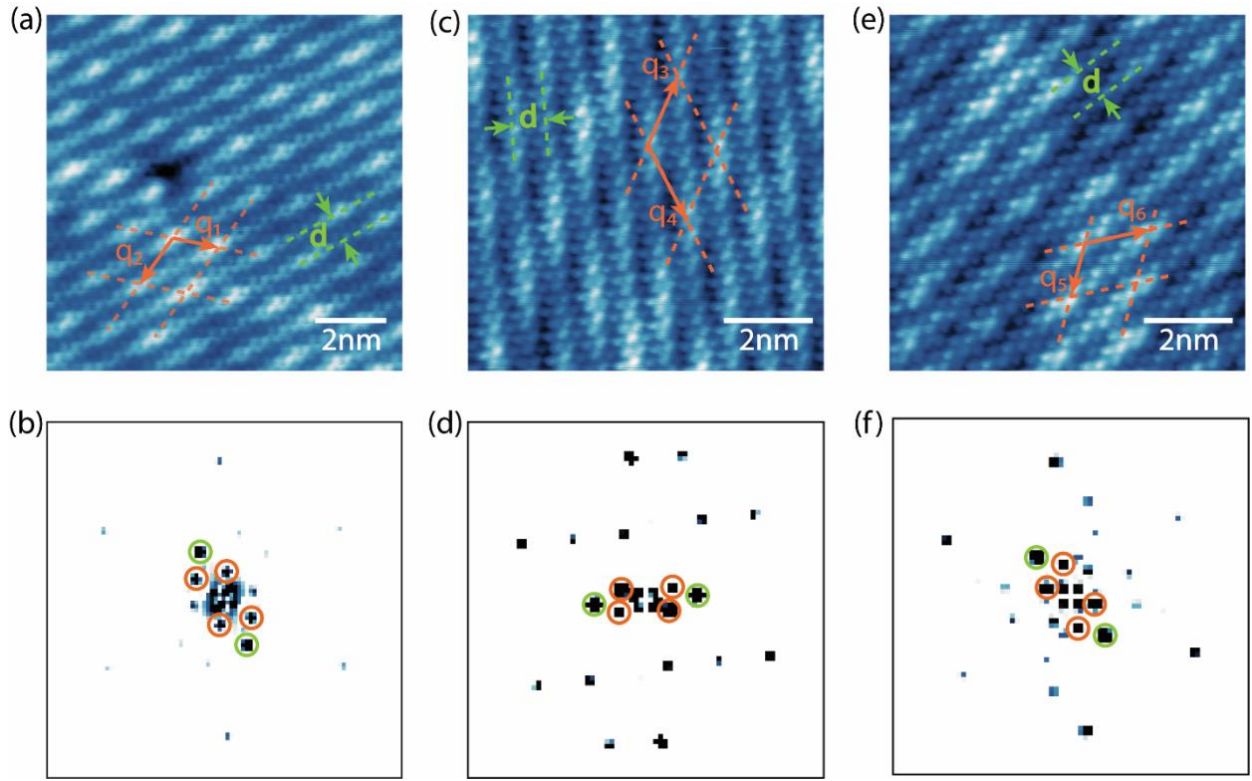

**Supplemental Figure 1| Different Charge Density Wave (CDW) patterns observed in the 1st layer of Low-Growth-Temperature (LGT) VSe$_2$ samples. a**, **c**, **e**, STM atomic resolution topography with different CDW patterns. CDW vectors are shown as orange arrows. The one-dimensional stripes, indicated by the dashed green lines, are clear in all figures and the distance between stripes is same as d = 2.33a. In **a**, $q_1 =$ 4.2a, $q_2 =$ 4.6a and the angle in-between is about 114 degrees. In **c**, $q_3 =$ 5.1a, $q_4 =$ 5.7a and the angle in-between is about 129 degrees. In **e**, $q_5 =$ 3.7a, $q_6 =$ 4.5a and the angle in-between is about 117 degrees. **b**, **d**, **f**, 2D Fast Fourier Transform (FFT) of figure **a**, **c**, **e**, correspondingly. The orange and green circles represent the 2D CDW order and 1D stripe order, respectively.

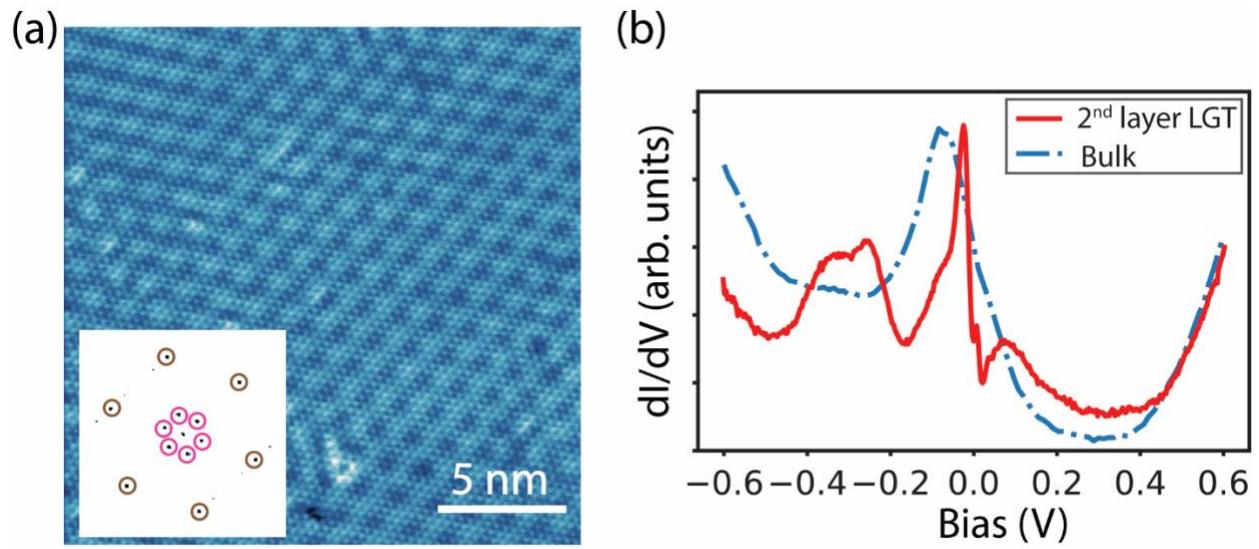

**Supplemental Figure 2| STM results on bulk VSe$_2$. a**, STM topography of freshly cleaved VSe$_2$ surface, in which 4a × 4a CDW is observed. The inset is the FFT of the topography. The brown circles represent the Bragg peaks and the pink circles indicate the CDW order. **b**, comparison of dI/dV spectra between the bulk and the 2$^{nd}$ layer of the LGT samples. Both pronounced peaks close to Fermi energy originate from V 3d band. The change of the peak position is because of the reduction of dimensionality.

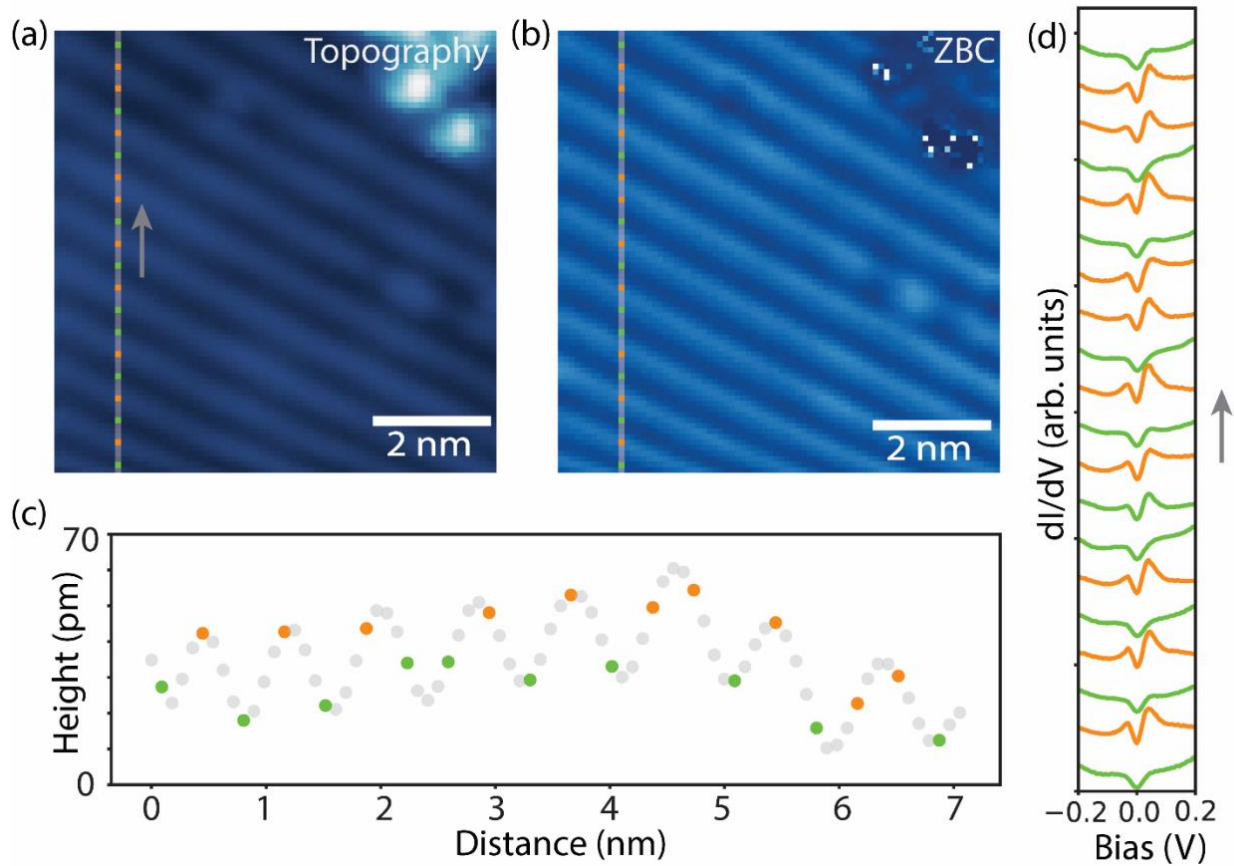

**Supplemental Figure 3| Comparison between of dI/dV spectra on bright stripes and dark stripes in the 1st layer of HGT samples. a**, topography obtained from a dI/dV map measured on 1st layer of an HGT sample, which shows a strong 1D feature. **b**, the zero-bias conductance (ZBC) image from the same dI/dV map. **c**, the height profile of the vertical linecut shown as a grey line in **a** with same color pattern. The orange dots are on bright stripes and green dots are on dark stripes. The grey dots represent the rest of the line. **d**, a waterfall-like plot of spectra (normalized to the conductance at -0.2 V respectively) on orange and green spots in **a** and **b**. It's clear there is an extra peak at 40 meV for all orange spectra, compared to the green ones. The difference between orange and green spectra is consistent.

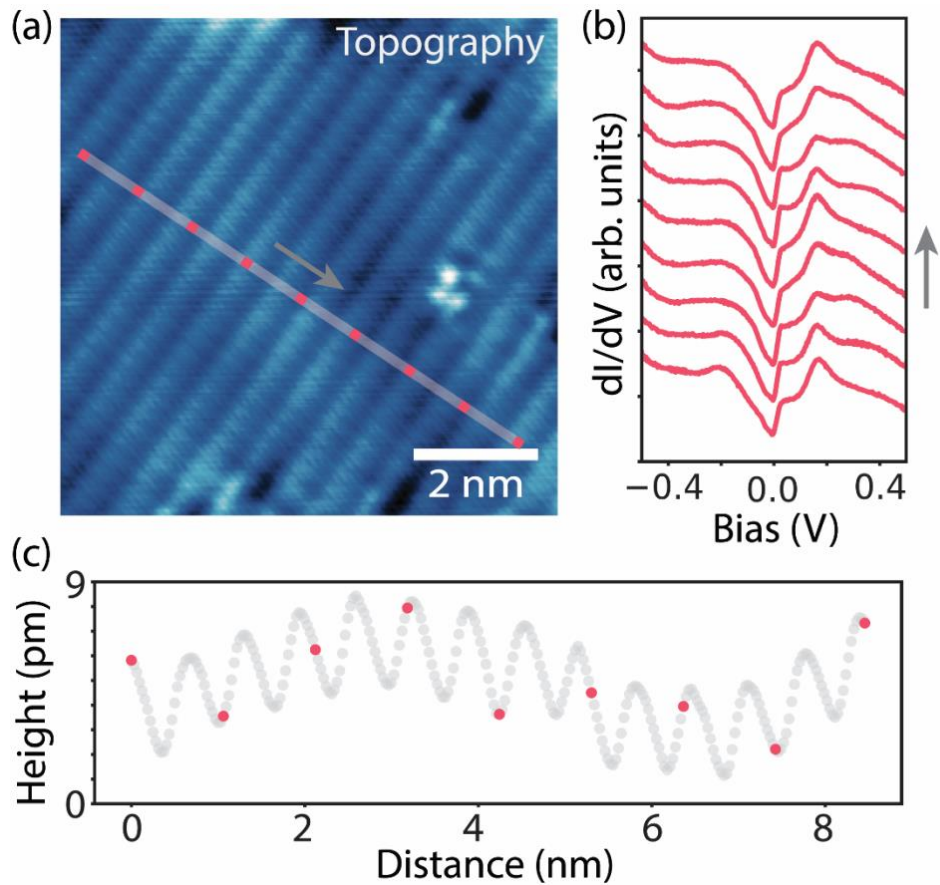

**Supplemental Figure 4| Comparison between of dI/dV spectra on bright stripes and dark stripes in the 2nd layer of HGT samples. a**, topography measured on 2nd layer of an HGT sample. **b**, a waterfall-like plot of spectra of the pink dots on the linecut shown in **a**. There is no obvious difference between the spectra on bright and dark stripes. It is consistent with the RHEED pattern which indicates the growth of distorted 1T-VSe$_2$. **c**, the height profile of the linecut shown as a grey line in **a** with same color pattern.

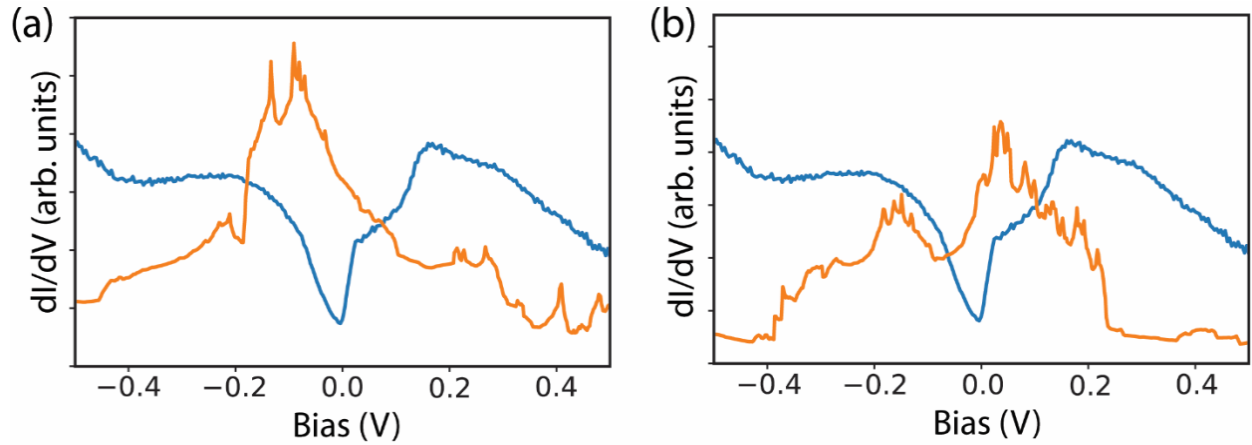

**Supplemental Figure 5| Comparison of dI/dV spectra between first-principles calculation result and experimental data. a**, calculated DOS curve for 1T' phase bilayer $VSe_2$ thin films (orange line) and DOS curve taken at the 2$^{nd}$ layer HGT samples (blue curves). **b**, calculated DOS curve for 1T$_d$ phase bilayer $VSe_2$ thin films (orange line) and DOS curve taken at the 2$^{nd}$ layer HGT samples (blue curves).

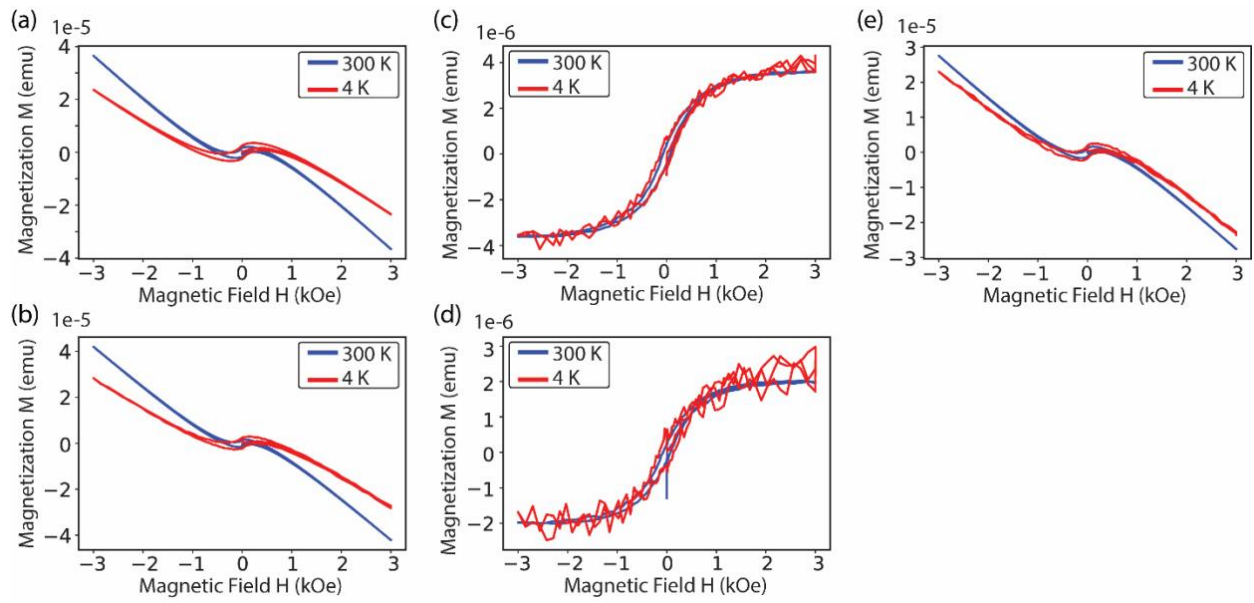

**Supplemental Figure 6| MPMS results of VSe$_2$ thin films and background. a**, **b**, M-H curves measured at 4 K and 300 K for 1-layer LGT VSe$_2$ sample and 1.5-layers HGT VSe$_2$ sample with in-plane magnetic field, respectively. The samples consist of Se-capped MBE-grown VSe$_2$ film on bilayer graphene on SiC wafer pieces. The diamagnetic background comes from the substrate. **c**, **d**, M-H curves in **a** and **b** after subtracting the linear background and the contribution from the control sample, which will be explained later in **e**. The hysteresis loops still exist even at 300 K for both samples. The saturation magnetization reads ~ $3 \times 10^{-6}$ emu and ~$2 \times 10^{-6}$ emu in **c** and **d**, which are equivalent to ~3.5 $\mu_B$ and ~1.3 $\mu_B$ per formula unit. **e**, M-H curve measured for the control sample at 4 K and 300 K. The control sample has same setup, including the same plastic sample holder and other necessary accessories, without any sample.